\documentclass[twocolumn,nofootinbib,showpacs,10pt]{revtex4-1}
\usepackage{color,graphicx}
\usepackage[colorlinks,hyperindex]{hyperref}

\definecolor{green1}{RGB}{0,128,0} 
\hypersetup{hidelinks,backref=true,pagebackref=true,hyperindex=true,colorlinks=true,breaklinks=true,urlcolor= blue}
\hypersetup{%
  colorlinks = true,
  linkcolor  = blue,
  citecolor = green1,
}
\usepackage{bookmark}
\usepackage{hyperref,color,xcolor}
\hypersetup{hidelinks,hyperindex=true,colorlinks=true,breaklinks=true,urlcolor= blue}
\hypersetup{%
  colorlinks = true,
  linkcolor  = blue
}

\usepackage{amssymb,amsmath,upgreek}
\usepackage{graphicx,float}
\usepackage{indentfirst}
\newcommand{\be}{\begin{equation}}
\newcommand{\ee}{\end{equation}}

\newcommand{\ba}{\begin{eqnarray}}
\newcommand{\ea}{\end{eqnarray}}

\usepackage{color}
\usepackage{tensor}
\usepackage{braket}
\usepackage{bm}

\newcommand{\de}{\mathrm{d}}

\newcommand{\hypF}{{}_2\mathrm{F}_1}

\begin{document}

\title{Extended quantum portrait of MGD black holes and information entropy}
\author{A. Fernandes--Silva}
\email{armando.silva@ufabc.edu.br}
\affiliation{CCNH, Universidade Federal do ABC - UFABC, 09210-580, Santo Andr\'e, Brazil.}

\author{A. J. Ferreira--Martins}
\email{andre.juan@ufabc.edu.br}
\affiliation{CCNH, Universidade Federal do ABC - UFABC, 09210-580, Santo Andr\'e, Brazil.}

\author{R. da Rocha}
\email{roldao.rocha@ufabc.edu.br}
\affiliation{CMCC, Universidade Federal do ABC -- UFABC, 09210-580, Santo Andr\'e, Brazil.}


\begin{abstract}The extended minimal geometric deformation (EMGD) 
is employed on the fluid membrane paradigm, to describe compact stellar objects as   
Bose--Einstein condensates (BEC) consisting of gravitons. The black hole quantum portrait, besides deriving a preciser phenomenological bound for the fluid brane tension, is then scrutinized from the point of view of the configurational entropy. It yields a range for the critical density of the EMGD BEC, whose  configurational entropy has global minima suggesting 
the configurational stability of the EMGD BEC. \end{abstract}
\maketitle
\flushbottom
%
%
%

\section{Introduction}

The AdS/CFT correspondence implements a non-perturbative representation of string theory, with appropriate  boundary conditions,  implementing holography. 
In particular, strongly coupled quantum field theories (QFTs)  
are well addressed in this framework.  The membrane paradigm can then be allocated in AdS/CFT, wherein black holes can be scrutinized 
 \cite{Casadio:2015gea,daRocha:2017cxu,Fernandes-Silva:2018abr,Cavalcanti:2016mbe}, in the long wavelength limit  \cite{Hubeny:2010wp,Ovalle:2019qyi}.  
Transport  coefficients,  measuring how \textcolor{black}{fast} a perturbed system returns to equilibrium, were introduced by the duality between   black branes in anti-de Sitter (AdS) space and hydrodynamics in its boundary \cite{Bhattacharyya:2008jc}. Perturbations can assemble the prototypical flow dynamics that underlies viscous fluids, modeled by the standard Navier--Stokes equations. Usually, the fluid is considered as living on  the border (the brane) of the AdS space (the bulk) 
The membrane paradigm consists of this approach, in the fluid/gravity  correspondence structure \cite{Shiromizu:2001jm,Shiromizu:2001ve}.

Mainly the bosonic sector of the fluid/gravity correspondence was used to study compact stellar distributions with the \textcolor{black}{minimal} geometric deformation method (MGD) \cite{Ovalle:2017fgl,Ovalle:2018vmg,ovalle2007,Casadio:2016aum}, where the relativistic elasticity emulates deformations of compact stellar objects   \cite{darkstars}. 
MGD black holes have been thoroughly investigated \cite{Casadio:2016aum,Casadio:2015gea,daRocha:2017cxu,Contreras:2018nfg,Contreras:2018gzd}, whose geometries  implement holographic dual objects to the boundary, low energy regime, fluid dynamics.   The MGD is a complementary framework to well-established setups \cite{Antoniadis:1998ig}.  
The main ingredient of the MGD is implemented by the 4D brane tension ($\upsigma$) \cite{Ovalle:2010zc,Ovalle:2013vna,covalle2}. It is capable to govern the high energy corrections to General Relativity (GR), which  is the MGD $\upsigma \rightarrow \infty$ low energy regime. As the 4D brane temperature has been remarkably altered throughout the universe evolution, a  variable tension fluid brane is a natural landscape to implement the MGD \cite{gly2,Casadio:2013uma,daRocha:2017cxu,Casadio:2016aum}. Fluid membranes can fluctuate due to small intrinsic (surface) 
tension. Describing the boundary of AdS/CFT as 
a fluid brane, analogously, the cosmological constant 
can also fluctuate. A constitutive part of an effective action, the cosmological constant is then emulated by the membrane 
free energy.  Similarly to the surface tension of a 2D fluid membrane presenting  statistical fluctuations \textcolor{black}{due to} its  molecular structure, its molecular length corresponds to the Planck length, in describing the 4D boundary as a fluid brane \cite{Casadio:2015gea,Abdalla:2009pg}. 
In this setup, the membranes statics correspond to the spacetime dynamics \cite{Samuel:2006ga,Katti:2009mi}. 

On the other hand, attempts of quantization of Einstein's classical GR  naturally lead to the massless spin-2 graviton, whose perturbative unitarity is violated in graviton-graviton scattering processes at distances much smaller than Planck's length. It, thus, precludes a perturbative description of gravity at such short scales. To address this issue, Ref. \cite{1112-1-2} argued that GR can be seen as a QFT in a self-complete manner, without the necessity of a Wilsonian UV-completion. This idea is denominated classicalization. From the quantum point of view, the classicality of a system is measured by $N$, the occupation number of bosons in the classical field of interest, so that a classical object is realized as \textcolor{black}{a} bound-state of high occupation number, $N \gg 1$ \cite{1112-2}. The quantum physics of black holes is fully given in terms of $N$, so that black hole semiclassical and geometric features such as thermality, entropy and horizon are emergent notions arising in the large-$N$ limit \cite{Dvali:2011aa}. Thus, in the quantum $N$-portrait of black holes, $N$ fully provides a quantum-mechanical description of a black hole, independently of classical geometric notions. Hence, a black hole is seen as a Bose--Einstein condensate (BEC) of $N$ weakly-interacting bosons, in a leaky bound-state accounting to Hawking radiation, realized through the quantum depletion process of the BEC.  In fact, in this picture, black holes are realized as critical phenomena, as, more precisely, they are at the brink of a quantum phase transition. The classicalization phenomena comes into play, as a black hole is the configuration in which $N$ is maximized, \textcolor{black}{and} it is the system whose classical wavelength corresponds to the Schwarzschild radius. In other words, for a given configuration of a characteristic wavelength, black holes will yield the most classical configuration, and a self-sustained system with maximally packed gravitons in the BEC configuration \cite{plb7,Dvali:2011aa,casadio_bec,Casadio:2015jha}. 

The main aim here is to take the extended MGD protocol (EMGD) 
to describe a black hole that is composed by a BEC of gravitons. Therefore, the Chandrasekhar-like  critical density and the BEC stability are  derived, using the configurational entropy (CE) apparatus. The CE has roots in the Shannon's  information entropy and has been an useful setup to probe 
stable configurations of compact fermionic and bosonic stellar distributions. 
It consists of 
an evaluation of the information that is necessary to describe any given system configuration \cite{glst,glsow,Gleiser:2018kbq,Sowinski:2015cfa,Bernardini:2016hvx,Bernardini:2016qit}. In the context of a BEC, the CE is a natural tool that takes into account the $N$ occupation number as constituting the system modes.  Compact distributions are also driven by critical points of the CE, underlying diverse physical systems \cite{Correa:2015vka,Braga:2016wzx,Lee:2017ero,Correa:2016pgr,Bazeia:2018uyg}. 

This paper is assembled as follows: after briefly introducing the EMGD protocol in Sect. \ref{sMGD}, the black hole quantum portrait paradigm is then adopted to describe the black hole as a  
BEC of gravitons, in Sect. \ref{sec:bhqp}. Based on the absence of additional physical singularities in the EMGD black hole, we derive a 
more precise phenomenological bound for the fluid brane tension. Sect. \ref{4123} is dedicated to scrutinize the CE that underlies the EMGD BEC, deriving a stellar critical  density that emulates the Chandrasekhar critical density, represented by a global minimum of the CE. 
In addition, for a specific value of the parameter driving the EMGD, the CE  
derives a value of  the BEC mass and the characteristic length scale of the condensate, below which the compact stellar distribution  
does not collapse, irrespectively of its density. Sect. \ref{V} is devoted to the concluding remarks.

\section{Extended minimal geometric deformation}
\label{sMGD}
The main challenge in theories that describe gravity \textcolor{black}{and} seek to extend General Relativity (GR), is to derive physically feasible solutions of the  effective field equations, whose low energy limit recovers the GR Einstein's  equations. The extended minimal geometric deformation (EMGD) consists of a protocol that addresses this problem, as presented in Ref. \cite{Casadio:2015gea}.
In the membrane paradigm, the universe we live in is implemented by a  4D  brane, with the  brane tension, $\upsigma$, playing the role of its energy density. The brane itself can be recast as the AdS boundary. The 4D effective  Einstein equations can be derived when one projects 5D field tensors onto the brane, by the Gauss--Codazzi method. It yields the following  effective energy-momentum tensor \cite{GCGR},
\begin{equation}
T^{eff}_{\mu\nu} = T_{\mu\nu} + \frac{6}{\upsigma} S_{\mu\nu} + \frac{1}{8\pi} \mathbb{E}_{\mu\nu},
\end{equation}
where $T_{\mu\nu}$ represents the brane stress-energy tensor, $S_{\mu\nu}$ \textcolor{black}{is} the second order stress-energy tensor, 
encrypting high energy corrections, and $\mathbb{E}_{\mu\nu}$ is the electric component of the Weyl tensor encoding the Weyl fluid interpretation, that trickles the bulk. Vacuum solutions, $T_{\mu\nu}=0$, yield the 4D effective Einstein's field equations,
\begin{equation}
G_{\mu\nu}=  \frac{\upkappa^2}{8\pi} \mathbb{E}_{\mu\nu},\label{einsteinvacuum}
\end{equation}
where $\upkappa^2/8\pi$ is Newton's coupling constant. 
It is worth to emphasize that the $\mathbb{E}_{\mu\nu}$  tensor has a non-local nature yielding purely gravitational effects from Kaluza-Klein bulk graviton fields modes. 
From the holographic point of the view, Refs.  \cite{Shiromizu:2001jm,Shiromizu:2001ve}
derived the brane CFT partition function,
\begin{equation}
Z  =  \int {\cal D}h\, e^{\frac{i}{2}S_{\rm b}} e^{iS_{\rm ct}}\Bigl\langle \exp \Big(i \int d^4x\; h_{\mu\nu}T^{\mu\nu}\Big) \Bigr\rangle_{\rm CFT} \ , \label{pathintegral}
\end{equation}
where, $h_{\mu\nu} = g_{\mu\nu} + u_\mu u_\nu$ is the induced metric, projecting quantities  onto the brane, the $u_\mu$ field represents a $4$-velocity field and $S_{\rm b}$ denotes the brane action, whereas  $S_{ct}$ is an usual counter-term, deploying a finite action.
From Eq. \eqref{pathintegral} one can obtain the effective Einstein's  equation on the brane. It is then possible to identify $\mathbb{E}_{\mu\nu}$ as the CFT stress-energy tensor. Thereby, the electric part of the Weyl tensor reduces to 
\begin{eqnarray}
\mathbb{E}_{\mu\nu} &=& -\frac{1}{8\pi}\langle T_{\mu\nu}
-\frac14 h_{\mu\nu}T\rangle_{\rm CFT}\nonumber\\&&\qquad  -\frac{1}{3}\left[D_\mu \phi D_\nu \phi
-\frac{1}{4}h_{\mu\nu}(D \phi)^2 \right], \label{Eads}
\end{eqnarray}
where $D_\mu$ denotes the covariant derivative with respect to the induced metric $h_{\mu\nu}$ and the matter fields on the brane are assumed, as usual, to couple with the scalar field $\phi$ which  plays the role of a dilaton in AdS/CFT. Therefore, the electric part of the Weyl tensor can be interpreted, in the AdS/CFT setup, as an outcome of the CFT energy-momentum tensor.
The first analysis that one can make about the Weyl tensor action is  the calculation of the effective pressure generated by it, for the static, spherically symmetric solution 
\begin{equation}
 ds^2 = -{\rm f}_1\mathrm{d}t^2 + {\rm f}_2\mathrm{d}r^2 + r^2\mathrm{d}\Omega^2 \ .\label{sphericalsy}
\end{equation}
The Weyl tensor can be split off as
\begin{equation}
k^{2}\,\mathbb{E}_{\mu \nu }=\frac{6}{\upsigma }\left[\mathbb{U}\left(u_{\mu }\,u_{\nu }+\frac{1}{3}\,h_{\mu \nu }\right)+\mathbb{P}_{\mu \nu }\right]\ , \label{weyltensor}
\end{equation}
where $\mathbb{U}$ is the bulk Weyl scalar and  $\mathbb{P}_{\mu \nu }$ is the stress tensor. 
By solving Eq. \eqref{einsteinvacuum}, using Eqs. \eqref{sphericalsy} and  \eqref{weyltensor}, it is possible to obtain the effective pressure components, 
\begin{equation}
{p}_{r}\,=\frac{\upsigma^{-1}}{3}\left(\mathbb{U}  +  2\mathbb{P}  \right)\   , \quad  \quad 
{p}_{t}\,= \frac{\upsigma^{-1}}{3}\left(\mathbb{U} -  \mathbb{P} \right),
\end{equation}
where $\mathbb{P}=\mathbb{P}_\mu^{\ \mu} $. Hence, one can measure the resultant anisotropy from the Weyl tensor by the difference between the components ${p}_{r}-{p}_{t}=\upsigma^{-1}{\mathbb{P}}$. The expected isotropy is then recovered in the GR limit. 

The EMGD procedure assumes that a deformation term is added into the radial and temporal components of \eqref{sphericalsy} \cite{Casadio:2015gea}, 
\begin{equation}
 {\rm f}_1(r) = 1-\frac{2\,M}{r}+f(r)  \ ,\quad  {\rm f}(r) _2 = 1-\frac{2\,M}{r} +h(r) \ , \label{defcomp}
\end{equation}
where $f(r)$ denotes the geometric deformation of the radial metric component and $h(r)$ the temporal deformation, accordingly.
Eq. \eqref{einsteinvacuum} and the traceless condition of the Weyl tensor, $\mathbb{E}^\mu_{\ \mu}=0$, implies that $R^\mu_{\ \mu}=0$, which, by using Eq. \eqref{defcomp}, yields
\begin{equation}
F(f)+H(h)=0 \label{fh} \ ,
\end{equation}
where, denoting ${\rm f}_1(r)=e^{\upnu(r)}$ and ${\rm f}_2(r)=e^{\lambda(r)}$, 
\begin{eqnarray}
\!\!\!\!\!\!\!\!\!\!\!\!\!\!\!\!\!\!\!\!F(f) \!&=\!&\! {\left(\frac{\upnu'}{2}\!+\!\frac{2}{r}\right)}\,f'\!+\!{\left(\upnu''+\frac{{\upnu'}^2}{2}\!+\!\frac{2\upnu'}{r}+\frac{2}{r^2}\right)}\,f \ ,
\\
\!\!\!\!\!H(h) &=& \lambda'\,\frac{h'}{2}+\lambda\left(h''\!+\!\upnu_s'\,h'+\frac{h'^2}{2}+2\,\frac{h'}{r}\right),
\end{eqnarray} and  the prime denotes derivatives with respect to the $r$ coordinate. It is interesting to notice that a constant $h$ implies $H=0$, which recovers the MGD case \cite{Casadio:2015gea,Ovalle:2016pwp}.
In addition, the most general solution can be obtained by solving Eq. \eqref{fh}. In this case, the exterior temporal metric component reduces to
\begin{equation}
{\rm f}_1(r) = \left ( 1 - \frac{2M}{r}\right )^{k+1},\label{generaltempo}
\end{equation}
where  $k$ is the deformation parameter. Additionally, the radial metric component reads \cite{Ovalle:2016pwp} 
\begin{eqnarray}
{\rm f}_2(r)&=&1-2\,M/r \nonumber \\  &&+ \frac{1}{r^k}(1-2\,M/r)^{(1-k)}\Big[(k-3)M+2r\Big]^{\frac{3-5k}{k-3}}\nonumber \\ &&\times\left(\frac{-2M}{r-2M}\right)^k\left(1\!+\!\frac{2r}{(k-3)M}\right)^{\frac{4k}{k-3}} \!\!\! \nonumber \\  &&\times\Big[\upsigma^{-1}r^{\frac{k(k+1)}{k\!-\!3}}\Big(1\!-\!\frac{r}{2\,M}\Big)^k  \Big(1+\frac{2r}{(k-3)M}\Big)^{\frac{4k}{k-3}}\nonumber \\ && -(k-3)M\,r^k(1-2\,M/r)\times \nonumber \\ &&\Big[(k-3)M+2r\Big]^{\frac{4k}{k-3}}\,F(r,k)\Big], \label{generalradial}
\end{eqnarray}
where $F(r,k)\!\!\!=\!\!\! F_1{\left[\frac{k(k\!+\!1)}{3-k}, 1\!-\!k,  \frac{4k}{3-k}, \frac{3 \!+\! k^2}{ 3 -k}; \frac{r}{2M}, \frac{2\,r}{(3-k)M}\right ]}$ stands for the standard Appell function.

Different solutions \textcolor{black}{have} been analyzed for different values of $k$. For $k=0$, the deformed exterior temporal and radial metric components are, respectively, given by the standard MGD  \cite{Casadio:2015gea},
\begin{eqnarray}
 \!\!\!\!\!\!\!\!\!\!{\rm f}_1(r) &=& 1 - \frac{2M}{r}, \label{k0tempo}
\\
 \!\!\!\!\!\!\!\!\!\!{\rm f}_2^{-1}(r) &=&  \left ( 1 \!+\! \frac{R\left(1-\frac{3M}{2R}\right)}{1-\frac{2M}{R}} \frac{\upsigma^{-1}}{r \left (1- \frac{3M}{2r} \right )}\right )\left (1 \!-\! \frac{2M}{r} \right )\label{k0radial}.
\end{eqnarray}
The radial component \eqref{k0radial} contains all the information about the deformation, that essentially depends on $\upsigma^{-1}$. 
Clearly in the GR limit,  the metric components \eqref{k0tempo} and \eqref{k0radial}  compel the gravitational field of the MGD solution, that is weaker when compared to the GR limit.  For the case when $k=1$,  the temporal and radial components  depict a Reissner--Nordstr\"om-like metric. 
As \textcolor{black}{seen for} $k=0$, the deformed term decreases the intensity of the gravitational field with respect to the Schwarzschild case. The event horizon for the Reissner--Nordstr\"om-like metric is then degenerated at  $r_h = 2M$. Hence, for $k=1$ the event horizon is \textcolor{black}{indistinguishable from} the GR limit case.
The solution for $k=2$ is given by 
\begin{equation}
 {\rm f}_1(r)  = 1 - 2A + B - \frac{2}{9}AB \ ,
\end{equation}
and
\begin{eqnarray}
 \!\!\!\!\!\!\!\!\!\! {\rm f}_2^{-1}(r)  &=& \left( 1 - \frac{2A}{3}\right)^{-1} \left[ \frac{4A}{3} \left ( 1 - \frac{A}{6} \right )^7\right. \nonumber\\&&\left.+ \frac{5}{4644864} B^4 +\frac{5}{82944}\left(8 - A\right)B^3\right.  \nonumber
  \\
&&\left.\!\!\!+  \frac{25}{1728}\left(6\! -\! A\right)  B^2 \!+\! \frac{5}{12}\left (2\! -\! A\right )B \!-\!\frac{4A}{3} \!+\! 1 \right], 
\end{eqnarray}
where $A\equiv A(r) = 3M/r$ and $ B\equiv B(r)= 12M^2/r^2$.
In this case, the degenerate event horizon is at $r_h \approxeq 3.36 M $, which is larger than the Schwarzschild case, $r_h=2M$. Therefore, AdS bulk effects increase the EMGD black hole horizon. \textcolor{black}{Thus}, for the analyzed cases, the intensity of the gravitational field is lower than the GR case, caused by the action of the Weyl fluid. 

The general-relativistic classical tests, taking into account an EMGD  sun, yield  $k \lesssim 4.2$ \cite{Fernandes-Silva:2018abr}. Hence, as the metric radial component (\ref{generalradial}) is not defined for $k=3$, the cases $k=1$ and $k=2$, and $k=4$ are studied in Sect. \ref{4123}.
\textcolor{black}{The $k = 4$ case is a relevant one, besides being a new exact exterior solution for a spherically
symmetric self-gravitating system \cite{Ovalle:2015nfa}. Terms of higher-order, with respect to the terms in the tidal charge, $Q$, define this solution, where the temporal metric (\ref{generaltempo}) is explicitly given by Eq. (30) in Ref. \cite{Ovalle:2015nfa}. The $k=4$  solution has a singularity, $r_s$, that is concealed behind the event horizon, ${\rm h}$, which is smaller than the standard Schwarzschild  event horizon. The decrement of the event horizon, compared to the Schwarzschild solution, is caused exclusively by the bulk Weyl fluid. The bulk Weyl scalar, $\mathbb{U}$, was shown to increase, along 
approximation towards the stellar distribution, up to a peak  
inside the horizon \cite{Ovalle:2015nfa}. It diverges
 afterward, as one reaches the $r_s$ singularity. In addition, 
 the trace of the stress tensor, $\mathbb{P}$, attains negative values, in the $k=4$ case. Moving away from the compact stellar distribution, 
both the bulk Weyl scalar and the trace of the stress tensor fastly vanish, suggesting a Weyl atmosphere enclosing the compact distribution. Besides, the Weyl fluid for the $k=4$ case is weaker than previously analyzed cases 
\cite{Casadio:2015gea}. }

\section{Black hole quantum portrait}
\label{sec:bhqp}
In this section the quantum portrait of a black hole, as a 
BEC of gravitons, is briefly 
reviewed and better analyzed. 
\subsection{The quantum model}

One can describe a black hole as a BEC of gravitons \cite{Muck:2014kea}, whose quantum mechanical behavior is captured by the solution of the $s$-wave modes of the time-independent Schr\"odinger equation for a particle of mass $m$ in a spherically symmetric P\"oschl-Teller potential, namely
\begin{equation}
      \left( -\frac{\hbar^2}{2m} \, \nabla^2 +V(r) \right) \uppsi(\bm{x}) = E \uppsi(\bm{x})~,
\label{eq:eds}
\end{equation}
\noindent with
\begin{equation}
      V(r)=-\upxi(\upxi+1)\frac{\hbar^2\upomega^2}{2m}\sec\!{\rm h}^2(\upomega r).
\label{eq:posch_pot}
\end{equation}
\noindent The constant $\upomega$ has units of inverse-length and sets the characteristic length scale of the system, whereas $\upxi>0$ is a dimensionless parameter to be fixed in the following discussion. 
It is assumed that the potential is generated by the condensate itself, so that its spherical symmetry is only reasonable if there is no angular momentum, which is why we discard the $l>0$ modes. Further, the $l=0$ modes represent the $s$-waves, which are the only components effectively taking part in the scattering process, which further supports the choice of focusing only in the $l=0$ modes. This being the case, by decomposing the wave function in terms of the spherical harmonics, $
\uppsi(\bm{x}) = \frac{1}{r}R(r)   Y_{lm}(\theta, \phi)$, 
one arrives at the radial Schr\"odinger equation with $l=0$ for $R(r)$:
\begin{equation}
\left( -\frac{\hbar^2}{2 m} \frac{\de^2}{\de \textcolor{black}{r}^2} + V(r) \right) R(r) = E R(r).
\label{eq:eds_radial}
\end{equation}
\noindent Its solution, restricted by the regularity condition of $\uppsi$ at the origin, is given in terms of the Gauss hypergeometric function ${}_2\mathrm{F}_1$ by \cite{Muck:2014kea} 
\begin{eqnarray}
R(r) &=&
\hypF \left(\alpha_++\frac12,\alpha_-+\frac12;\frac32;-\sinh^2(\upomega r) \right)\nonumber\\&&\times\,
{\sinh(\upomega r)}{\sec\!{\rm h}^\upxi(\upomega r)}\,a,
\label{eq:radial_sol}
\end{eqnarray}
\noindent where $a$ is a normalization constant and
\begin{equation}
\alpha_\pm = -\frac12 \left(\pm \upxi + i \frac{{\rm k}}{\upomega} \right) \ ,
\end{equation}
\noindent where the wave number $k$ is defined as the standard ${\rm k}^2 = 2mE/\hbar^2$.

For the bound states (for which $E<0$), the normalization condition imposes a discretization to the values of ${\rm k}$, given by \cite{Muck:2014kea} ${\rm k}_n = i \upomega \left ( \upxi - 1 -2n \right )$, for $n=0,1,\ldots, \left[\frac{\upxi-1}2\right]$. It characterizes the discrete bound states. On the other hand, the states with $E>0$ form a continuum of scattered states, which are asymptotically given by a superposition of incoming and outgoing spherical waves \cite{Muck:2014kea}. 
This description yields the full spectrum and the eigenstates of the proposed Hamiltonian, given in terms of the independent parameters $\upxi, m$ and $\upomega$. Without further constraints, it might lead to superluminal scattering states, whose lack of physical interpretation compels one to set a relativistic quantum mechanical model. This apparatus can be achieved if one now takes the field $\uppsi$ to be a relativistic complex Klein--Gordon field, $\uppsi(t, \bm{x}) \equiv \uppsi(x)$, which is a solution to the Klein--Gordon equation (one considers only positive energies) with rest mass $\mu$ and two independent potentials, $S(\bm{x})$ (scalar) and $V(\bm{x})$ (vector): 
\begin{equation}
\left( \hbar^2 \nabla^2 \!+\! \left(i\hbar\partial_t \!-\!V(\bm{x})\right)^2 \!-\! \left(S(\bm{x}) \!+\! \mu\right)^2 \right) \uppsi(x) =0 \ .
\end{equation}

The potentials are time independent, which allows us to write for the energy $\epsilon$, $\uppsi(x) = \mathrm{e}^{-i\epsilon t/\hbar} \uppsi(\bm{x})$. Thus, the particular choice $V=S$ yields
\begin{equation}
\left(-\frac{\hbar^2}{2(\epsilon+\mu)} \nabla^2 + V \right ) \uppsi(x) = \frac12( \epsilon-\mu) \uppsi(x)~,
\end{equation}
\noindent which is just Eq. \eqref{eq:eds} with constraints $
m= \epsilon+\mu$ and $E= \frac12 \left(\epsilon-\mu\right)$. Hence, by choosing $V=S=V(r)$, the P\"oschl-Teller potential of Eq. \eqref{eq:posch_pot}, one can directly identify the solution of Eq. \eqref{eq:radial_sol} as the radial solution to the relativistic model --- but subjected to the constraints, which, together with our previous definition of the wave number, $\hbar^2{\rm k}^2 = 2mE$, represent the relativistic dispersion relation, $
\epsilon^2 - \mu^2 = \hbar^2 {\rm k}^2$. 
One now follows Refs. \cite{casadio_bec,Muck:2014kea} and impose the so-called marginal binding condition, which fixes the ground state (of energy $\epsilon=0$) as the single bound state, whilst the first excited state is identified with the onset of the continuum. This condition realizes the Hawking radiation in the BEC black hole model through the quantum depletion process, as an excited graviton is emitted from the BEC as a scattering state. 

These conditions lead to important constraints. First, one fix  the graviton effective mass as $\mu = \hbar \upomega$. As $\epsilon_{n=1} = \epsilon_{{\rm k}=0} = \mu$, this means that the energy gap between the ground state and the continuum is precisely the graviton effective mass. In addition, for the ground state ($\epsilon = n = 0$), one has $\upxi = 2$, which fixes the P\"oschl-Teller potential as 
\begin{equation}
V(r) = {-3\hbar \upomega}\,{\sec\!{\rm h}^2(\upomega r)} \ .
\label{eq:posch_grd}
\end{equation}
The  discussion above makes clear that in the relativistic setting the only free parameter consists of  the scale $\upomega$, which is identified to the graviton's effective de Broglie wavelength. As shall be seen later on, the most natural value for this parameter within the quantum portrait scenario \cite{Dvali:2011aa,casadio_bec,Casadio:2015jha} leads us to the interpretation of the graviton BEC as a black hole. 

Before  proceeding, the norm of the ground state reads 
\begin{equation}
\begin{aligned}
\braket{\uppsi_0 | \uppsi_0} &= -\frac2{\hbar} \int_{\mathbb{R}^3}\de^3 x\, V |\uppsi_0|^2 =
\\
&=6a^2 \int\limits_0^\infty \de u\, {\sinh^2 u}\,{\sec\!{\rm h}^6 u} =  \frac{4a^2}{5} \ ,
\end{aligned}
\end{equation}
\noindent which fixes $a^2=5/4$. In the above integral, $u \equiv \upomega r$\footnote{We used the fact that at the ground state we have $\upxi_0 = 2$ as well as $\alpha_+=-1/2$ and $\alpha_-=-3/2$, so that the hypergeometric function of the radial solution in \eqref{eq:radial_sol} equals 1. Thus, $R(r) = a \frac{\sinh(\upomega r)}{\cosh^2(\upomega r)}$. Afterwards, since $|\uppsi_0|^2 = \frac{a^2}{r^2}\frac{\sinh^2(\upomega r)}{\cosh^4(\upomega r)}\frac{1}{4\pi}$, in order to compute the first integral one only has to substitute $V(r)$ as in Eq. \eqref{eq:posch_grd}, and $\de^3 x = 4\pi r^2 \de r$, to derive the second integral with $u = \upomega r$.}.

\subsection{EMGD BEC black hole}

After constructing the quantum mechanical model, now the graviton BEC is scrutinized from the macroscopic point of view. It can be modelled as an anisotropic fluid, whose local stress tensor takes the form
\begin{equation}
\textcolor{black}{T^{\mu\nu} = (p_\parallel-p_\perp) v^\mu v^\nu +p_\perp g^{\mu\nu} + (p_\perp + \upepsilon) u^\mu u^\nu} \ ,\label{000}
\end{equation}
\noindent where the  $u^\mu$ and $v^\mu$ vector fields are unit, orthogonal, and respectively time-like and space-like. The energy density is denoted by $\upepsilon$ and the $p_\perp$ and $p_\parallel$ denote, respectively, the pressures that are orthogonal and parallel to the $v^\mu$ field.

As discussed in Sect. \ref{sMGD}, the EMGD method is implemented upon the metric (\ref{sphericalsy}). Hence, 
\begin{equation}
u^\mu = \left(1/\sqrt{{\rm f}_1(r)},0,0,0\right)^\intercal, \;\;\; v^\mu = \left( 0, 1/\sqrt{{\rm f}_2(r)}, 0, 0\right)^\intercal \ .
\end{equation}
Now, the EMGD is a method employed to solve the effective Einstein field equations on the brane, Eq. (\ref{einsteinvacuum}), so that  the interior region is governed by Eq. \eqref{generaltempo} \cite{Casadio:2015gea}:
\begin{equation}
{\rm f}_1(r)= \left ( 1 - \frac{2 m(r)}{r} \right )^{k+1} \ ,
\label{eq:enu}
\end{equation}
\noindent where 
$m(r) = 4\pi \int_0^r \bar{r}^2 \upepsilon(r) \de \bar{r}$ 
represents the quasilocal Misner--Sharp mass function, appearing in the interior solution instead of $M$ as in Eq. (\ref{generaltempo}). 

Now, the main point provided by the quantum portrait of black holes paradigm, which allows one to interpret a black holes as a  BEC of gravitons, comes into play when the fluid energy density $\upepsilon$ (which characterizes the BEC macroscopically) is identified to the charge density of the ground state complex Klein--Gordon field $\uppsi(x)$ (which represents the BEC at the microscopical level). By doing this,  the Klein--Gordon charge can be thus  interpreted as a gravitational charge, which allows one to make the connection between a graviton BEC and a black hole. With this quite important identification, the Misner--Sharp mass function reads
\begin{equation}
\begin{aligned}
m(r) 
&=\frac{M}2 \tanh^3(\upomega r) \left[ 5 -3 \tanh^2(\upomega r) \right] \ .
\label{bh:M.expl}
\end{aligned}
\end{equation} 
In fact, one explicitly accounts that the squared norm of the complex Klein--Gordon field yields the gravitational mass $m(r)$, as one interprets $\upepsilon$ as its charge. This is how one deploys the bridge between the microscopic BEC description and the gravitational description of the EMGD black hole through its metric, by substituting Eq. \eqref{bh:M.expl} into Eq. \eqref{eq:enu} and rewriting it as:
\begin{equation}
\begin{gathered}
 \!\!\!\!\!\!{\rm f}_1^\vartheta(\rho) = \left (1 - \frac{1}{\rho} \tanh^3 (\vartheta \rho) \left (5 - 3 \tanh^2 (\vartheta \rho) \right ) \right )^{k+1}\!,
\end{gathered}
\end{equation}
\noindent where $\vartheta \equiv M \upomega $ and $\rho \equiv \frac{r}{M}$.
It is worth to realize that, as made explicitly clear by the added \textcolor{black}{superscript}, ${\rm f}_1^\vartheta(\rho)$, now depends on the parameter $\vartheta$, which expresses nothing but the graviton de Broglie wavelength $\upomega$, scaled by the total mass of the BEC $M$. Given this, an immediate question is whether there exists any interval for the parameter $\vartheta$, limiting the existence of the black hole. Here one shall address this question by analyzing the sign of the function ${\rm f}_1^\vartheta(\rho)$, since a horizon will exist if it is possible to have ${\rm f}_1^\vartheta(\rho)\leq 0$. Now, the derivative of $f_1^\vartheta(\rho)$ with respect to $\rho$  reads 
\begin{widetext}
 \begin{equation}
\begin{gathered}
\!\!\!\!\!\frac{\de {\rm f}_1^\vartheta(\rho)}{\de \rho} = \left(\frac{\rho\!-\!5 \tanh^3(\vartheta \rho)\!+\! 3\tanh^5(\vartheta \rho)}{\rho}\right)^k \left (\frac{(1\!+\!k) \, \text{sech}^4(\vartheta \rho) \left[8 \sinh(2 \vartheta \rho)\!-\!60 \vartheta \rho\!+\! \sinh(4 \vartheta \rho) \right] \tanh^2(\vartheta \rho)}{4 \rho^2} \right ) \ .
\end{gathered}
\end{equation}
\end{widetext}
This derivative equals zero at the point of local minimum $y_{\text{min}} =  \vartheta \rho_{\text{min}}$, such that $
 60 y_{\text{min}} = 8 \sinh (2 y_{\text{min}}) + \sinh(4y_{\text{min}})$,  which has the numerical solution $y_{\text{min}} \approx 1.03121$. Now, if one rewrites ${\rm f}_1^\vartheta(\rho)$ evaluated at this point of minimum as
\begin{equation}
\begin{aligned}
\!\!\!\!\!\!{\rm f}_1^\vartheta(\rho_{\text{min}}) &= 1 - \frac{\vartheta}{y_{\text{min}}} \left [ \tanh^3 (y_{\text{min}}) \left (5 - 3 \tanh^2 (y_{\text{min}}) \right ) \right ]
 \\
 &\equiv 1 - \frac{\vartheta}{\vartheta_{\text{min}}},
\end{aligned}
\end{equation} 
\noindent then this only characterizes a horizon if $ \vartheta \geq \vartheta_{\text{min}}$. Written as above, it is straightforward to see that
\begin{equation}
\vartheta_{\text{min}} = \frac{y_{\text{min}}}{\tanh^3 (y_{\text{min}}) \left (5 \!-\! 3 \tanh^2 (y_{\text{min}}) \right )} \approx 0.69372.
\end{equation}
Hence, in order for the black hole to exist, one must have $
 \vartheta \gtrsim 0.69372$. 
As argued in Ref. \cite{Muck:2014kea}, the most natural value for $\vartheta$ following the BEC quantum portrait of black holes is $\vartheta = 1 > \vartheta_{\text{min}}$, so that in fact one can have a BEC black hole. It then becomes clear that, indeed, the black hole depends on the BEC parameters.

\begin{figure}[t]
\includegraphics[scale=0.4]{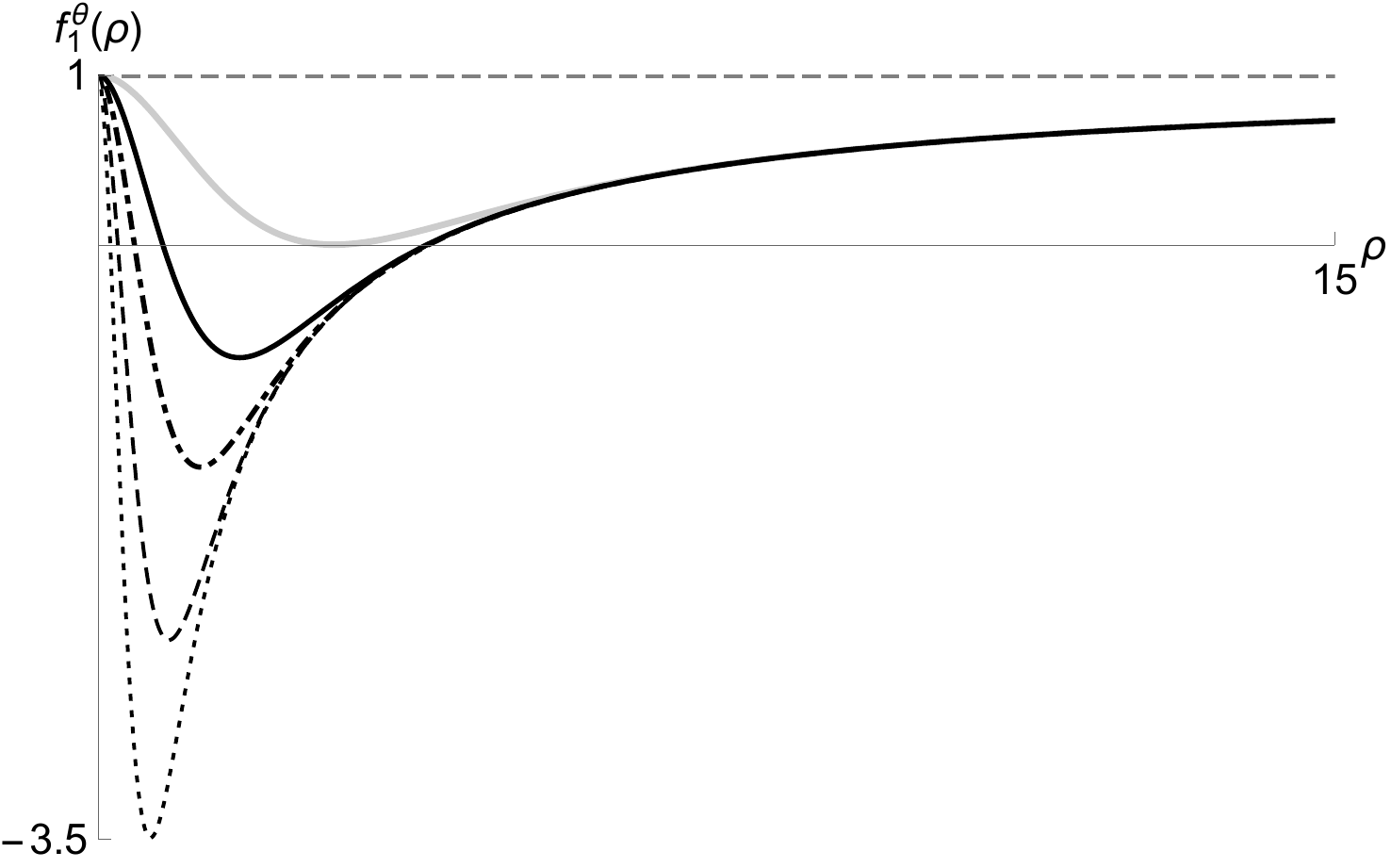}
\caption{Plot of ${\rm f}_1^\vartheta(\rho)$ in Eq.~\eqref{Brho},
for $\vartheta = 0$ (dashed grey line);
$\vartheta = 0.3$ (grey thick line);
$\vartheta = 0.5$ (black thick line);
$\vartheta = \vartheta_{\rm min}$ (black dot-dashed line);
$\vartheta = 1$ (black dashed line);
$\vartheta = 1.35$ (dotted line). 
}\label{fig1}
\end{figure}

Assuming the EMGD setup, the radial metric, to leading order in
$\upsigma^{-1}$ reads
\begin{equation}
\frac{{\rm f}_1^\vartheta(\rho)}{{\rm f}_1(\rho)}
=
1-
\frac{b}{\upsigma\big\{\!\left[\rho-\frac{3}{4} \tanh^3(\vartheta \rho)\right]
\!\!\left[ 5-3\tanh^2(\vartheta\rho)\right]\big\}^{k+1}}
\ ,
\label{Brho}
\end{equation}
where ${b}\simeq 0.55$.
Fig.~\ref{fig1} shows plots of ${\rm f}_1^\vartheta(\rho)$ for various values of $\vartheta$.
It is clear that, for increasing values of $\vartheta$, this black hole model
rapidly approaches the Schwarzschild black hole.
\par
Now, this setup can be better analyzed in the  fluid membrane 
paradigm. \textcolor{black}{In fact, the brane tension constancy, assumed in the original Randall-Sundrum 1 model can be emulated by a setup with an effective stress-tensor on the brane. In fact, generalizations of the original Randall-Sundrum scenario were implemented in Refs. \cite{gly1,gly2,maar} on  Friedmann branes.  Also,  Ref. \cite{european} derived an explicit expression for the brane tension  that implements cosmological inflation, then reheating, a deceleration period and also late time acceleration, precisely matching the CMB anisotropy revealed by WMAP. 
It has observational evidence for an inflationary paradigm. Variable tension branes, with analogy to fluid membranes, manifest a temperature-dependence governed by the E\"otv\"os law. It permits running gravitational and cosmological ``constants''.}

\textcolor{black}{In an inflationary model,} the brane tension varies as the temperature decreases when the Universe inflated  \cite{Abdalla:2009pg}. This can be usually implemented by the so-called E\"otv\"os law, ruling the brane tension as a linear function of the temperature  \cite{gly2,Abdalla:2009pg}.  Another way to realize this dependence is to 
\textcolor{black}{recall} that the surface tension is proportional 
to the membrane temperature, $T$. In fact,  a membrane in equilibrium may vibrate due to thermal fluctuations with respect to its equilibrium configuration. Small vibrations can be then ruled by harmonic
oscillators, wherein the expectation value of energy in each mode
is $T$. Performing a sum over all modes yields the  brane tension \cite{Samuel:2006ga} $
\upsigma
\propto 
\left(T_{\rm crit}-T\right),$ 
where  $T_{\rm crit}$ denotes a critical temperature.
The tension variation is now expressed in terms of the (cosmological) time,
instead of the temperature. The CMB indicates $T\sim a^{-1}$, where $a=a(t)$ is the  universe scale parameter~\cite{gly2}, $
\upsigma(t)=\upsigma_0\left(1-\frac{a_0}{a(t)}\right)$, where 
and $a_0$ denotes a critical.
\par
\begin{figure}[H]
\begin{center}
\includegraphics[scale=0.59]{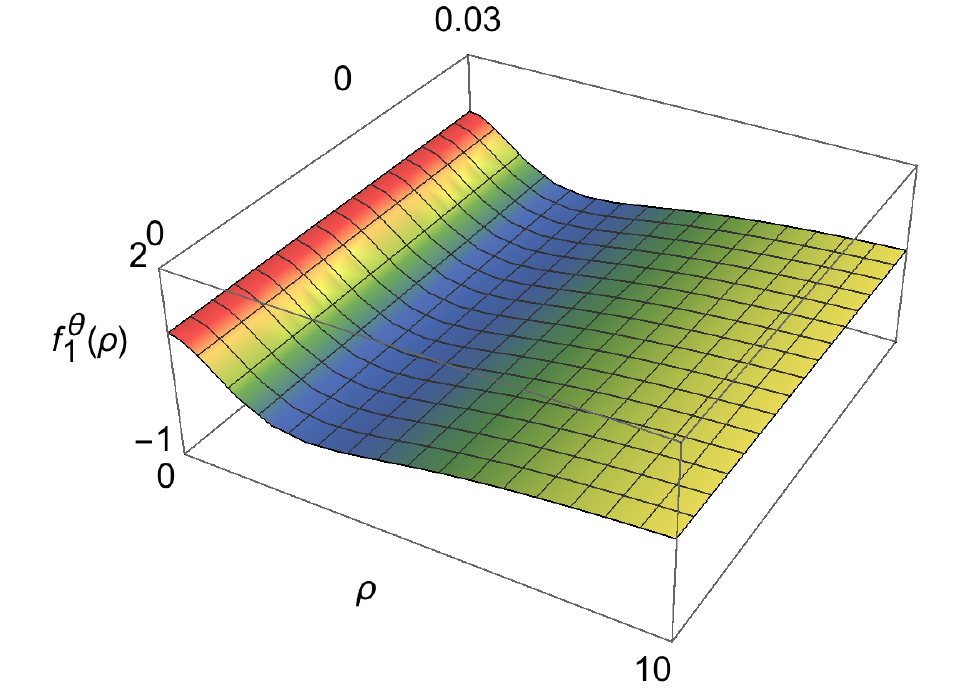}
\includegraphics[scale=0.57]{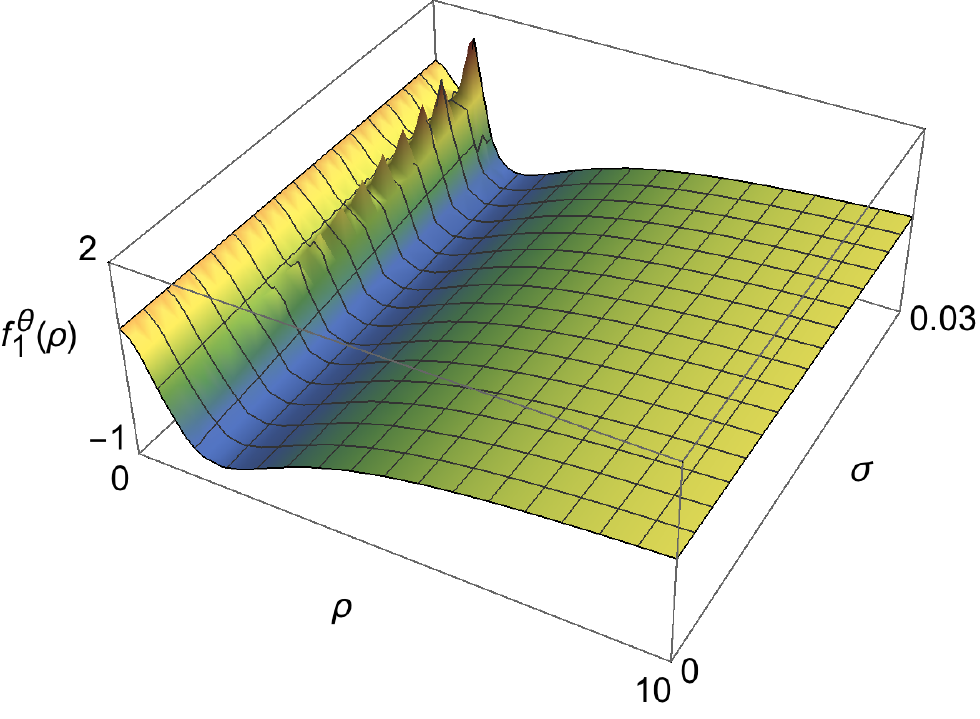}
\includegraphics[scale=0.57]{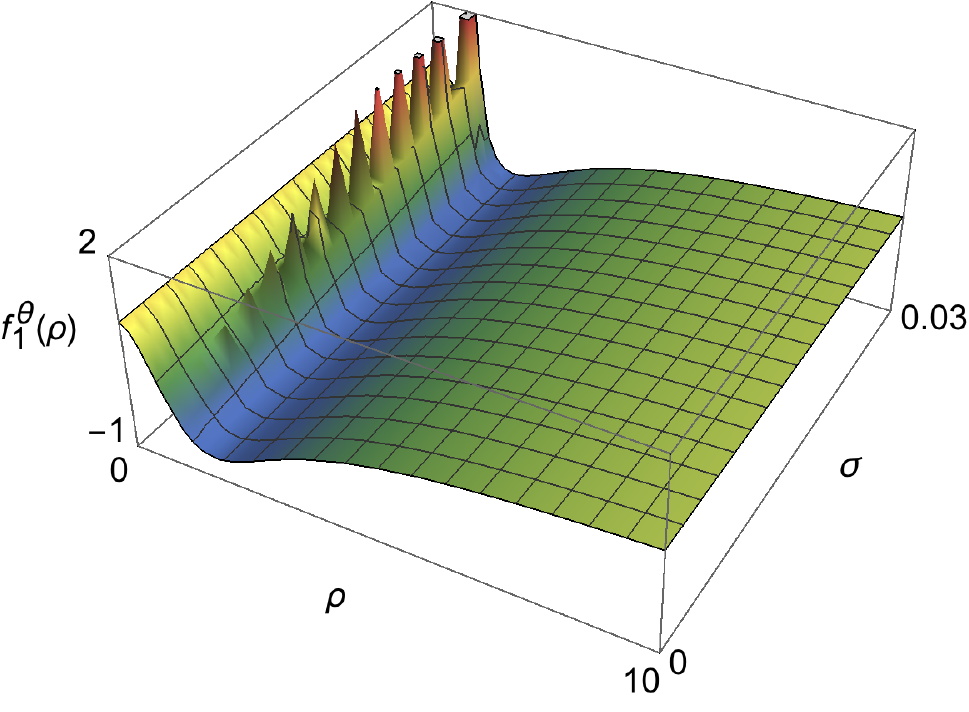}
\caption{Plot of ${\rm f}_1^\vartheta(\rho)$ for $\vartheta = 0.3$,  for $k=0$ (upper panel), $k=1$ (middle panel) and $k=2$ (bottom panel),  in the era dominated by the cosmological constant,
as a function of brane tension $\upsigma(t)$ and the radial coordinate $\rho={r}/{M}$.}
 \label{F40}
\end{center}
\end{figure}\begin{figure}[H]
\begin{center}
\includegraphics[scale=0.59]{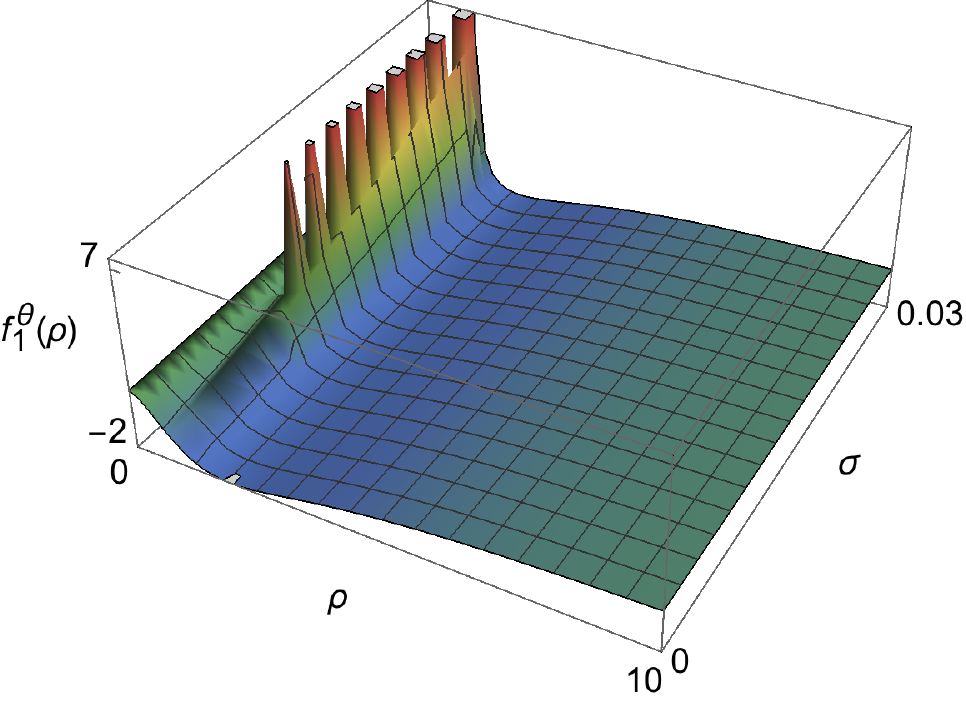}
\caption{Plot of ${\rm f}_1^\vartheta(\rho)$ for $\vartheta = 0.3$,  for  $k=4$, in the era dominated by the cosmological constant,
as a function of brane tension $\upsigma(t)$ and the radial coordinate $\rho={r}/{M}$.}
 \label{F4}
\end{center}
\end{figure}
\par
{The plots in Figs.  \ref{F40} and \ref{F4} indicate that additional singularities might exist. However, computing the scalar invariants $R$, $R_{\mu\nu}R^{\mu\nu}$ and  $R_{\mu\nu\rho\sigma}R^{\mu\nu\rho\sigma}$ for the EMGD metric (\ref{generaltempo}, \ref{generalradial}), for the $k=1$, $k=2$, and $k=4$ cases, no further 
singularities are obtained if the brane tension is, respectively, bounded by 
\begin{equation}
\upsigma\gtrapprox\begin{cases}& 2.94 \times 10^6 {\rm MeV}^4, \;\;\text{for $k=1$},\\
&2.91 \times 10^6 {\rm MeV}^4, \;\;\text{for $k=2$},\\
&2.83 \times 10^6 {\rm MeV}^4, \;\;\text{for $k=4$}
\ .\end{cases}
\end{equation}
{This phenomenological bound, thus, refines the previous  limit $\upsigma\gtrapprox 3.18 \times 10^6 {\rm MeV}^4$ \cite{Casadio:2016aum}.}
\section{Configurational entropy and information stability of EMGD BEC}
\label{IV}
\label{4123}
The Fourier transform 
$T_{00}(k) = \int_\mathbb{R}T_{00}(r)e^{-ik\cdot r}\,dr,$ with respect to the radial coordinate implements the so called modal fraction  $\upzeta(k) = \frac{|T_{00}(k)|^{2}}{ \int_{\mathbb{R}}  |T_{00}(k)|^{2}dk}.$  It is a correlation probability distribution that quantifies how much a $k$ wave  mode contributes to the power spectrum, associated to the energy density.
The CE, then, measures the information content of the spatial profile that characterizes the energy density, $T_{00}(r)$, with respect to the Fourier wave modes. The CE 
 is defined by \cite{Gleiser:2012tu,Sowinski:2015cfa}
\begin{eqnarray}
S_c = - \int_{\mathbb{R}}\mathring\upzeta(k)\log {\mathring\upzeta}(k)\, dk\,,
\label{confige}
\end{eqnarray}
where $\mathring\upzeta(k)=\upzeta(k)/\upzeta_{\rm max}(k)$.
\begin{figure}[H]
\begin{center}
\includegraphics[width=2.6in]{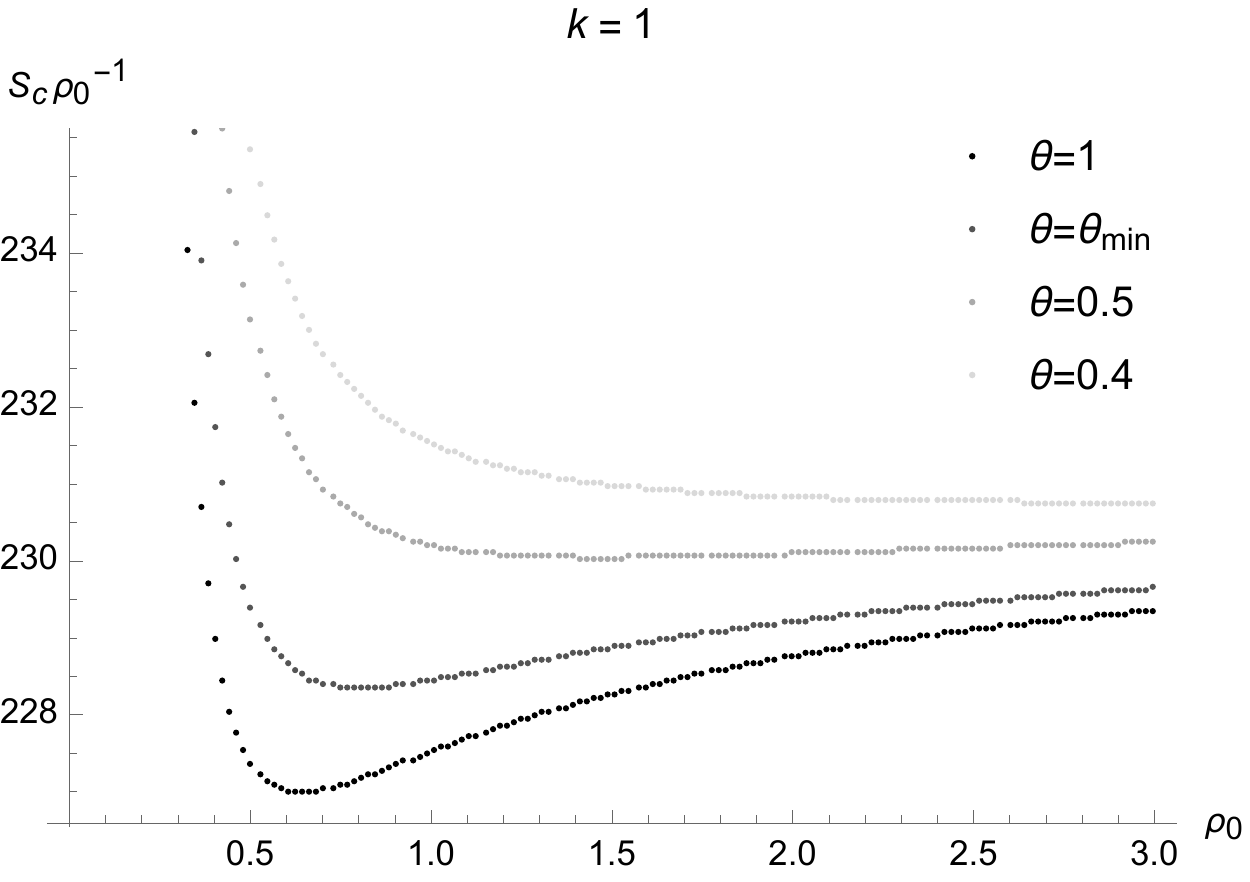}
\includegraphics[width=2.6in]{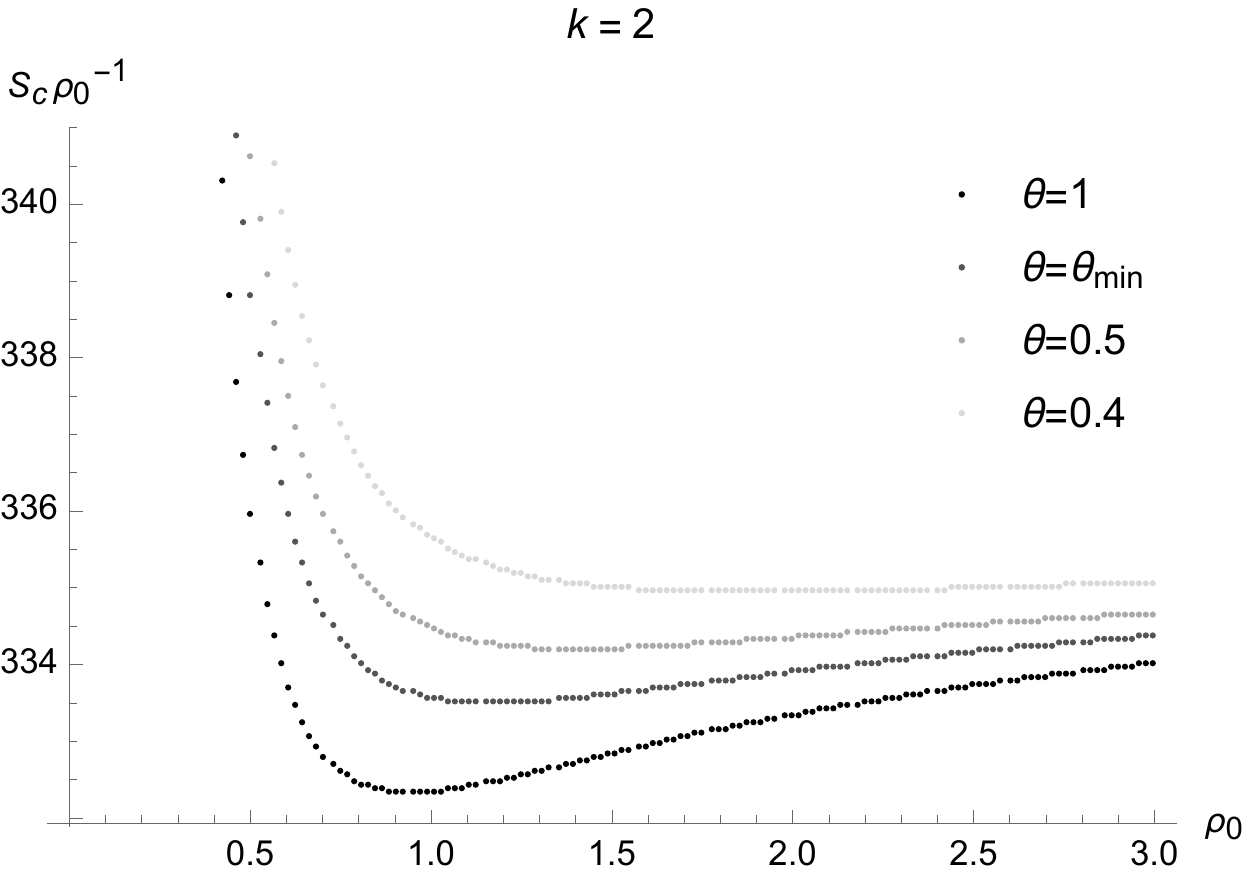}
\includegraphics[width=2.6in]{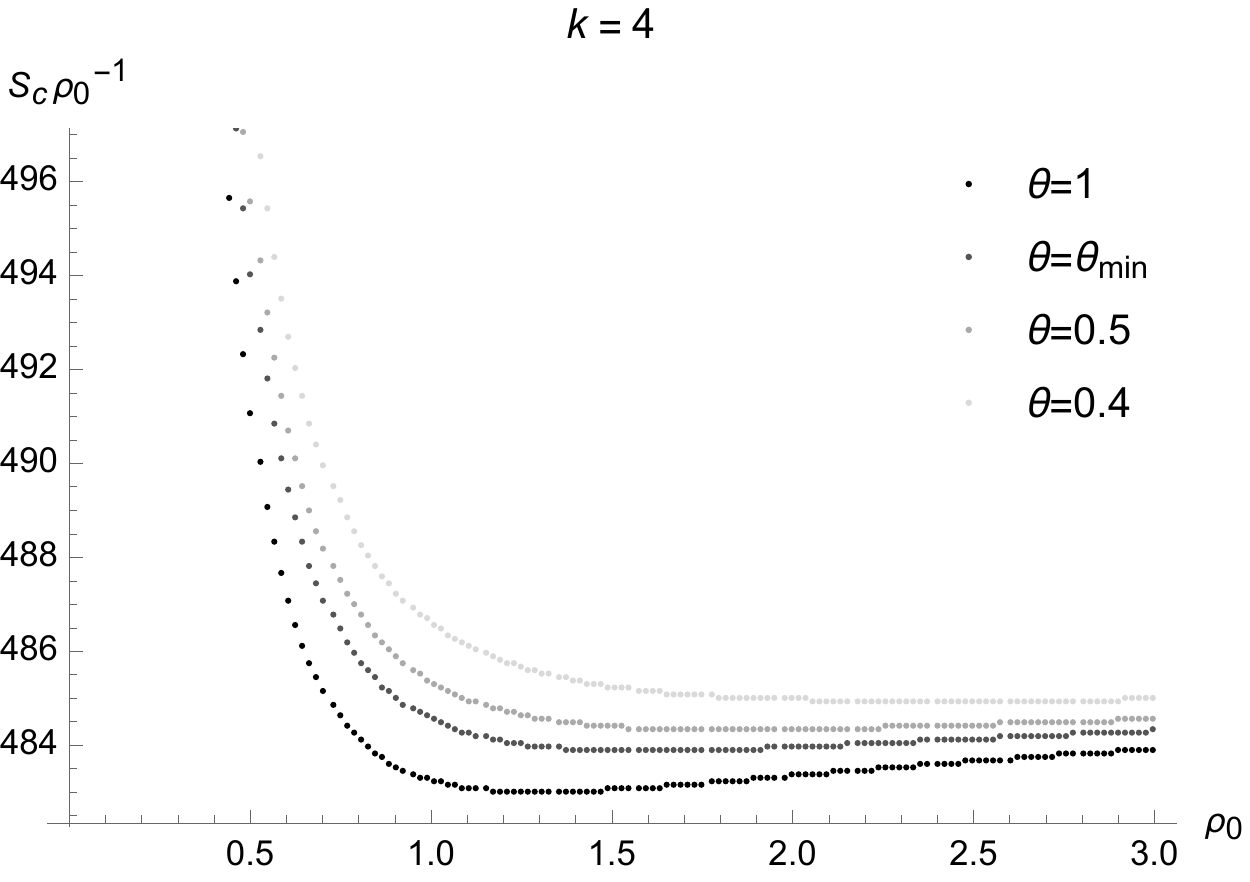}
\caption{Configurational entropy as a function of the inverse of the BEC EMGD black hole critical  density $\rho_0$, for $k=1$, $k=2$ and $k=4$, with respect to $\rho_0$.}
\label{f3}
\end{center}
\end{figure}
The equilibrium configurations can be derived in terms of the energy density, obtained by the $T_{00}$ component in Eq. (\ref{000}).
Then the CE, $S_c$, is plot  as a function of the critical central density  of the EMGD BEC compact stellar distribution, denoted by  
$T_{00}(0)= \rho_0$.
The CE \eqref{confige} is then computed and shown in Fig.~\ref{f3}. The CE is multiplied by the inverse central density to produce a quantity
that scales with dimensions of inverse mass. 
\par
The analysis of the CE for compact distributions 
makes use of the fiducial value, $\rho_c$, of the critical central density.  
In Refs.~\cite{Gleiser:2015rwa,Casadio:2016aum}, the critical points of the CE 
were related to $\rho_c$ and to the Chandrasekhar density as well.
The plots in Fig.~ \ref{f3} also emulate and extend previous results in the context of the MGD setup \cite{Casadio:2016aum}. 
The less information involved in the modes that comprise the EMGD BEC,  the less entropic information is required to represent its configuration. In the upper plot in Fig.~\ref{f3}, which represents the EMGD BEC for $k=1$, 
there is a critical value  $\vartheta_{\rm crit.}\approxeq 0.13$ below which the CE has no global minimum. Then in the CE context, for value of $\vartheta\lesssim \vartheta_{\rm crit.}$, such kind of compact stellar distributions 
are less likely to collapse, irrespectively of their $\rho_0$ density. It does not occur for any value of $\vartheta$ in the $k=2$ and $k=4$ cases. 
In addition,  the higher the value of $k$, the higher the value of $S_c\rho_0^{-1}$ is. It is also worth to emphasize that Figs. \ref{f3} show 
another relevant pattern of the EMGD BEC. In fact, for a given fixed value of $\vartheta$, the higher the value of $k$, the higher the $\rho_0$ density  is, at global minima of $S_c\rho_0^{-1}$. For example, taking $\vartheta = 1$, 
the minimum of $S_c\rho_0^{-1}$ occurs at $\rho_0\approxeq 0.64$ for $k=1$; 
at $\rho_0\approxeq 0.92$ for $k=2$; and at $\rho_0\approxeq 1.27$ for $k=4$. 
In the next section, important features of the EMGD BEC compact  distributions ar further discussed. %
\section{Concluding Remarks}
\label{V}
The extended quantum portrait of black holes here studied 
established a EMGD BEC compact stellar distribution in the membrane paradigm. The plots in Figs. \ref{F40} and \ref{F4} indicate possible coordinate singularities, besides the standard Schwarzschild ones. The scalar invariants were then computed for the EMGD metric for all integer possible values of $k$ and no further 
singularities were observed if the brane tension obeys the bound $
\upsigma\gtrapprox2.83 \times 10^6 {\rm MeV}^4$. 
This value enhances the range for the fluid brane tension, surpassing the previous one $\upsigma\gtrsim 3.18 \times 10^6 {\rm MeV}^4$, in Ref.~\cite{Casadio:2016aum}. Then the CE showed itself a relevant tool 
for better probing the EMGD BEC. The CE global minima, for each fixed value of $k$ in the plots of Fig. \ref{f3}, vary with respect to  
the value of the parameter $\vartheta$ involving the EMGD BEC mass and the characteristic length scale of the BEC. For a fixed value of $k$, 
the BEC critical central  density, which corresponds to a global minimum of the CE, varies with respect to $\vartheta$. Indeed, the higher the value of $\vartheta$, the lower the BEC critical central  density is. It is in agreement 
with the interpretation of the critical central  density as a point of configurational stability of the EMGD BEC compact stellar distribution. 
Specifically for the $k=1$ case, the CE indicates that there is a value for the  EMGD BEC mass and its characteristic length scale, encoded by the $\vartheta$ parameter, below which the EMGD BEC  does not collapse, being configurationally stable. 
\section*{Acknowledgments}
The work of AFS was financed in part by the Coordena\c c\~ao de Aperfei\c coamento de Pessoal de N\'ivel Superior - Brasil (CAPES). 
\textcolor{black}{AJFM} is grateful to FAPESP (Grants No.  2017/13046-0  and No.  2018/00570-5) and to CAPES. 
RdR~thanks to FAPESP (Grant No.  2017/18897-8) and to the National Council for Scientific and Technological Development  -- CNPq (Grant No. 303293/2015-2), for partial financial support.


\begin{thebibliography}{0}%
\makeatletter
\providecommand \@ifxundefined [1]{%
 \@ifx{#1\undefined}
}%
\providecommand \@ifnum [1]{%
 \ifnum #1\expandafter \@firstoftwo
 \else \expandafter \@secondoftwo
 \fi
}%
\providecommand \@ifx [1]{%
 \ifx #1\expandafter \@firstoftwo
 \else \expandafter \@secondoftwo
 \fi
}%
\providecommand \natexlab [1]{#1}%
\providecommand \enquote  [1]{``#1''}%
\providecommand \bibnamefont  [1]{#1}%
\providecommand \bibfnamefont [1]{#1}%
\providecommand \citenamefont [1]{#1}%
\providecommand \href@noop [0]{\@secondoftwo}%
\providecommand \href [0]{\begingroup \@sanitize@url \@href}%
\providecommand \@href[1]{\@@startlink{#1}\@@href}%
\providecommand \@@href[1]{\endgroup#1\@@endlink}%
\providecommand \@sanitize@url [0]{\catcode `\\12\catcode `\$12\catcode
  `\&12\catcode `\#12\catcode `\^12\catcode `\_12\catcode `\%12\relax}%
\providecommand \@@startlink[1]{}%
\providecommand \@@endlink[0]{}%
\providecommand \url  [0]{\begingroup\@sanitize@url \@url }%
\providecommand \@url [1]{\endgroup\@href {#1}{\urlprefix }}%
\providecommand \urlprefix  [0]{URL }%
\providecommand \Eprint [0]{\href }%
\providecommand \doibase [0]{http://dx.doi.org/}%
\providecommand \selectlanguage [0]{\@gobble}%
\providecommand \bibinfo  [0]{\@secondoftwo}%
\providecommand \bibfield  [0]{\@secondoftwo}%
\providecommand \translation [1]{[#1]}%
\providecommand \BibitemOpen [0]{}%
\providecommand \bibitemStop [0]{}%
\providecommand \bibitemNoStop [0]{.\EOS\space}%
\providecommand \EOS [0]{\spacefactor3000\relax}%
\providecommand \BibitemShut  [1]{\csname bibitem#1\endcsname}%
\let\auto@bib@innerbib\@empty
\end{thebibliography}%


\begin{thebibliography}{99}
	
	\bibitem{Casadio:2015gea} R.~Casadio, J.~Ovalle and R.~da Rocha,
	Class.\ Quant.\ Grav.\  {\bf 32} (2015)  215020
	[arXiv:1503.02873 [gr-qc]].
	
	\bibitem{daRocha:2017cxu} R.~da Rocha,
	Phys.\ Rev.\ D {\bf 95}  (2017) 124017 
	[arXiv:1701.00761 [hep-ph]].
	
	\bibitem{Fernandes-Silva:2018abr} A.~Fernandes-Silva, A.~J.~Ferreira-Martins and R.~da Rocha,
	Eur.\ Phys.\ J.\ C {\bf 78} (2018)  631
	[arXiv:1803.03336 [hep-th]].
	
	\bibitem{Cavalcanti:2016mbe} R.~T.~Cavalcanti, A.~G.~da Silva and R.~da Rocha,
	Class.\ Quant.\ Grav.\  {\bf 33} (2016)  215007
	[arXiv:1605.01271 [gr-qc]].
	
	\bibitem{Hubeny:2010wp} V.~E.~Hubeny,
	Class.\ Quant.\ Grav.\  {\bf 28} (2011) 114007
	[arXiv:1011.4948 [gr-qc]].
	
	\bibitem{Ovalle:2019qyi} J.~Ovalle,
	Phys.\ Lett.\ B {\bf 788} (2019) 213
	[arXiv:1812.03000 [gr-qc]].
	
	\bibitem{Bhattacharyya:2008jc} S.~Bhattacharyya, V.~E.~Hubeny, S.~Minwalla and M.~Rangamani,
	JHEP {\bf 0802} (2008) 045 
	[arXiv:0712.2456 [hep-th]].
	
	\bibitem{Shiromizu:2001jm} T.~Shiromizu and D.~Ida,
	Phys.\ Rev.\ D {\bf 64} (2001) 044015
	[hep-th/0102035].
	
	\bibitem{Shiromizu:2001ve} T.~Shiromizu, T.~Torii and D.~Ida,
	JHEP {\bf 0203} (2002) 007
	[hep-th/0105256].
	
	\bibitem{Ovalle:2017fgl} J.~Ovalle,
	Phys.\ Rev.\ D {\bf 95} (2017) no.10,  104019
	[arXiv:1704.05899 [gr-qc]].
	
	\bibitem{Ovalle:2018vmg} J.~Ovalle and A.~Sotomayor,
	Eur.\ Phys.\ J.\ Plus {\bf 133} (2018) no.10,  428
	[arXiv:1811.01300 [gr-qc]].
	
	\bibitem{ovalle2007} J. Ovalle,
	Int. J. Mod. Phys. D {\bf 18} (2009) 837 [arXiv:0809.3547  [gr-qc]].%
	\bibitem{Casadio:2016aum} R.~Casadio and R.~da Rocha,
	Phys.\ Lett.\ B {\bf 763} (2016) 434
	[arXiv:1610.01572 [hep-th]].

	\bibitem{darkstars} J.~Ovalle, L. A.~Gergely and R.~Casadio, 
	Class. Quant. Grav. {\bf 32} (2015) 045015 [arXiv:1405.0252 [gr-qc]].
	
	\bibitem{Contreras:2018nfg} 
  E.~Contreras and P.~Bargue\~no,
  Eur.\ Phys.\ J.\ C {\bf 78}, no. 12, 985 (2018)
  [arXiv:1809.09820 [gr-qc]].
  
  \bibitem{Contreras:2018gzd} 
  E.~Contreras,
  Eur.\ Phys.\ J.\ C {\bf 78}, no. 8, 678 (2018)
  [arXiv:1807.03252 [gr-qc]].
	
	
		
	\bibitem{Antoniadis:1998ig} I.~Antoniadis, N.~Arkani-Hamed, S.~Dimopoulos and G.~R.~Dvali,
	Phys.\ Lett.\ B {\bf 436} (1998) 257 [arXiv:hep-ph/9804398].
	
		
	\bibitem{Ovalle:2010zc} J.~Ovalle,
	Mod.\ Phys.\ Lett.\ A {\bf 25} (2010) 3323 [arXiv:1009.3674 [gr-qc]].
	
	\bibitem{Ovalle:2013vna} J.~Ovalle, F.~Linares, A.~Pasqua and A.~Sotomayor,
	Class.\ Quant.\ Grav.\  {\bf 30} (2013) 175019 [arXiv:1304.5995 [gr-qc]].
	
	\bibitem{covalle2} R.~Casadio and J.~Ovalle,
	Gen. Relat. Grav. {\bf 46} (2014) 1669 [{arXiv:1212.0409 [gr-qc]}]. 
	%
	%
	
	
	
	\bibitem{gly2} L. A. Gergely, {Phys. Rev. D} {\bf  79} (2009) 086007 [arXiv: 0806.4006 [gr-qc]].
	



	
	
	\bibitem{Casadio:2013uma} R.~Casadio, J.~Ovalle and R.~da Rocha,
	Class.\ Quant.\ Grav.\  {\bf 31} (2014) 045016.
	[arXiv:1310.5853 [gr-qc]].
	
	\bibitem{Abdalla:2009pg} M.~C.~B.~Abdalla, J.~M.~Hoff da Silva and R.~da Rocha,
	Phys.\ Rev.\ D {\bf 80} (2009) 046003 [arXiv:0907.1321 [hep-th]].
	
	\bibitem{Samuel:2006ga} J.~Samuel and S.~Sinha,
	Phys.\ Rev.\ Lett.\  {\bf 97} (2006) 161302
	[cond-mat/0603804].
	
		\bibitem{Katti:2009mi} R.~Katti, J.~Samuel and S.~Sinha,
	Class.\ Quant.\ Grav.\  {\bf 26} (2009) 135018 
	[arXiv:0904.1057 [gr-qc]].
	
	\bibitem{1112-1-2} G.~Dvali, S.~Folkerts and C.~Germani,
	Phys.\ Rev.\ D {\bf 84} (2011) 024039 
	[arXiv:1006.0984 [hep-th]].
	
	\bibitem{1112-3} G.~Dvali, G.~F.~Giudice, C.~Gomez and A.~Kehagias,
	JHEP {\bf 1108} (2011) 108 
	[arXiv:1010.1415 [hep-ph]].
	
	\bibitem{1112-2} G.~Dvali, C.~Gomez and A.~Kehagias,
	JHEP {\bf 1111} (2011) 070 
	[arXiv:1103.5963 [hep-th]].
	
	\bibitem{Dvali:2011aa} G.~Dvali and C.~Gomez, 
	{
		Fortsch. Phys.} {\bf 61} (2013) 742 [arXiv:1112.3359 [hep-th]].
	
	\bibitem{plb7} R.~Casadio, A.~Giugno, O.~Micu and A.~Orlandi,
	Entropy {\bf 17} (2015) 6893 [arXiv: 1511.01279 [gr-qc]].
	
	\bibitem{casadio_bec} R.~Casadio and A.~Orlandi,
	JHEP  {\bf 1308} (2013) 025 
	[arXiv:1302.7138 [gr-qc]].
	
	\bibitem{Casadio:2015jha} R.~Casadio, R.~T.~Cavalcanti, A.~Giugno and J.~Mureika,
	Phys.\ Lett.\ B {\bf 760} (2016) 36 \textcolor{black}{[arXiv:1509.09317 [gr-qc]].}
	
	\bibitem{glst} { M. Gleiser and N.
		Stamatopoulos, Phys. Lett. B \textbf{713} (2012) 304 [arXiv:1111.5597 [hep-th]]. }
	
	\bibitem{glsow} { M. Gleiser and D. Sowinski,
		Phys. Lett. B \textbf{727} (2013) 272 [arXiv:1307.0530 [hep-th]]. }
	
	\bibitem{Gleiser:2018kbq} M.~Gleiser, M.~Stephens and D.~Sowinski,
	Phys.\ Rev.\ D {\bf 97}  (2018) 096007 
	[{arXiv:1803.08550 [hep-th]}].
	
	\bibitem{Sowinski:2015cfa} M. Gleiser and D. Sowinski, Phys.\ Lett.\ B {\bf 747} (2015) 125  [{arXiv:1501.06800 [cond-mat.stat-mech]}].
	
	\bibitem{Bernardini:2016hvx} A.~E.~Bernardini and R.~da Rocha,
	Phys. Lett. B {\bf 762} (2016) 107 [arXiv:1605.00294 [hep-th]].
	
	\bibitem{Bernardini:2016qit} 
  A.~E.~Bernardini, N.~R.~F.~Braga and R.~da Rocha,
  Phys.\ Lett.\ B {\bf 765} (2017) 81 
  [arXiv:1609.01258 [hep-th]].
	
	\bibitem{Correa:2015vka}
  R.~A.~C.~Correa and R.~da Rocha,
  Eur.\ Phys.\ J.\ C {\bf 75} (2015) no.11,  522
  [arXiv:1502.02283 [hep-th]].
	
	\bibitem{Braga:2016wzx} N.~R.~F.~Braga and R.~da Rocha,
	Phys.\ Lett.\ B {\bf 767} (2017) 386 [{arXiv:1612.03289 [hep-th]}].
	
	\bibitem{Lee:2017ero} C.~O.~Lee,
	Phys.\ Lett.\ B {\bf 772} (2017) 471  
	[arXiv:1705.09047 [gr-qc]].
	
	\bibitem{Correa:2016pgr} R.~A.~C.~Correa, D.~M.~Dantas, C.~A.~S.~Almeida and R.~da Rocha,
	Phys.\ Lett.\ B {\bf 755} (2016) 358 [arXiv:1601.00076 [hep-th]].
	
	\bibitem{Bazeia:2018uyg}
  D.~Bazeia, D.~C.~Moreira and E.~I.~B.~Rodrigues,
  J. Magn. Magn. Mater. {\bf 475} (2019) 734 \textcolor{black}{[arXiv:1812.04950 [cond-mat.mes-hall]].} 
	
	\bibitem{GCGR} T. Shiromizu, K. Maeda and  M. Sasaki,
	Phys. Rev. D {\bf 62} (2000) 024012 [{arXiv:gr-qc/9910076}].
	
	\bibitem{Ovalle:2016pwp} J.~Ovalle, R.~Casadio, A.~Sotomayor,
	Adv.\ High Energy Phys.\  {\bf 2017} (2017) 9756914  
	[arXiv:1612.07926 [gr-qc]].
	
	\bibitem{Ovalle:2015nfa}
  J.~Ovalle,
  Int.\ J.\ Mod.\ Phys.\ Conf.\ Ser.\  {\bf 41} (2016) 1660132
  [arXiv:1510.00855 [gr-qc]].
	
	\bibitem{Muck:2014kea} W.~M\"uck and G.~Pozzo,
	JHEP {\bf 1405} 128 (2014) [arXiv:1403.1422 [hep-th]].
	%
	%
	%
		\bibitem{gly1} L. A. Gergely, 
Phys. Rev. D {\bf  78} (2008) 084006 [arXiv:0806.3857 [gr-qc]].

\bibitem{maar} R. Maartens, Phys. Rev. D {\bf 62} (2000) 084023  [arXiv:hep-th/0004166]. 
	

\bibitem{european}  K. C. Wong, K. S. Cheng, T. Harko, 
{Eur. Phys. J. C} {\bf 68} (2010) 241 [arXiv:1005.3101 [gr-qc]].
	\bibitem{Gleiser:2012tu} M. Gleiser and N. Stamatopoulos, Phys.\ Rev.\ D {\bf 86} (2012) 045004 [{arXiv:1205.3061 [hep-th]}].
	
	\bibitem{Gleiser:2015rwa} M.~Gleiser and N.~Jiang,
	Phys.\ Rev.\ D {\bf 92} (2015) 044046 [arXiv:1506.05722 [gr-qc]].
	\end{thebibliography}
\end{document}